\documentclass{aastex}
\usepackage{amsmath}
\usepackage{graphicx}
\usepackage{amsfonts}
\usepackage{amssymb}%
\usepackage{enumerate}
\RequirePackage{color}

\begin{document}

\title{Corrected Newtonian potentials in the two-body problem with applications}
\shorttitle{Corrected Newtonian potentials}
\shortauthors{Anisiu et al.}

\author{M.-C. Anisiu\altaffilmark{1}} \and \author{I. Sz\"{u}cs-Csillik\altaffilmark{2}}
\affil{Romanian Academy, Tiberiu Popoviciu Institute of Numerical Analysis, Cluj-Napoca}
\and
\affil{Romanian Academy, Institute of Astronomy, Astronomical Observatory Cluj-Napoca}

\begin{abstract}
The paper deals with an analytical study of various corrected Newtonian
potentials. We offer a complete description of the corrected potentials, for
the entire range of the parameters involved. These parameters can be fixed for different models in order to obtain a good concordance with known data.
Some of the potentials are generated by continued fractions, and another one is derived from the Newtonian potential by adding a logarithmic correction. 
The zonal potential, which models the motion of a satellite moving in the equatorial
plane of the Earth, is also considered. The range of the parameters for which
the potentials behave or not similarly to the Newtonian one is pointed out.
The shape of the potentials is displayed for all the significant cases, as well as the orbit of {\it Raduga-1M 2} satellite in the field generated by the continued fractional potential $U_3$, and then by the zonal one. For the continued fractional potential $U_{2}$ we study the basic problem of the existence and linear stability of circular orbits. We prove that such orbits exist and are linearly stable. This qualitative study offers the possibility to choose the adequate potential, either for modeling the motion of planets or satellites, or to explain some phenomena at galactic scale.
\end{abstract}

\keywords{Celestial Mechanics; Newtonian potential; Two-body problem; Circular orbits}

\section{Introduction}

The idea of modifying the original $1/r$\ Newtonian potential starts with
Newton himself. In his \emph{Principia} he has already proposed a potential of
the form $A/r+B/r^{2}$\ and studied the relative orbit in this case too.
Potentials of this type have been physically justified later by Maneff (also
spelled Manev), in a series of papers starting with 
\cite{maneff1924}. 
A deep insight in the Maneff field can be found in \cite{diacu2000}.

In this paper we deal with an analytical study of some corrected Newtonian
potentials. The study is motivated by the fact that nowadays many authors
consider various corrected Newtonian potentials without being concerned
whether those potentials have or have not the properties of the Newtonian one.

It is known that the Newtonian potential $\mu/r$, regarded as a function of $r$, satisfies the following conditions:
\begin{enumerate}[(i)]
\item $\lim_{r \rightarrow 0+}=+\infty$;
\item $\lim_{r \rightarrow + \infty}=0$;
\item it decreases from $+\infty$ to $0$ as $r \in (0, +\infty)$.
\end{enumerate}
Its graph is the dashed one in Fig. 3.

We offer a complete description of some corrected potentials, for the entire
range of the parameters. All the corrected potentials are central and inhomogeneous; their role
is important, for example, in the inverse problem of dynamics \citep{bozis1997, anisiu2007}. 

We consider potentials derived recently by \cite{abd2014} from continued fractions, as well as the logarithmically corrected Newtonian potential introduced by \cite{mucket1977}.
We identify those which satisfy the conditions (i)-(iii) (at least for a certain range of the parameters), as the Newtonian potential does. These potentials are suitable to be used to explain some phenomena in the motion of satellites or planets.

As an application we choose to model the orbits of {\it Raduga-1M 2} satellite (launch date 28 January 2010) using the corrected Newtonian potentials. 
{\it Raduga-1M} satellites are military communication satellites, and are the geostationary component of the Integrated Satellite Communication System, where they work in conjuction with the highly eccentric orbit {\it Meridian} satellites.
  
When a relative orbit is designed using a very simple orbit model, then the control station of the formation will need to continuously compensate for the modeling errors and burn fuel. This fuel consumption, depending on the modeling errors, could drastically reduce the lifetime of the spacecraft formation. Using the corrected potentials can reduce the modeling errors \citep{szucs2013}.

Some potentials, as the logarithmically corrected one, do not satisfy at least one of the conditions (i)-(iii), and they can be used to model the motion at galactic scale. For example, such a potential was considered in cosmologies which avoid to involve dark matter \citep{kinney2001}, 
or to study the rotation curves of spiral galaxies \citep{fabries2009}.

Exponentially corrected potentials \citep{seeliger1895} are also of interest.

A further study will be dedicated to the restricted three-body problem, and
regularization methods will be applied for a better understanding of the
motion, as in \cite{roman2014}.

Section 2 introduces the potentials generated by continued fractions. It
starts with some theoretical results, which allow us to establish a clear
difference between the odd and even such potentials. A special attention is
given to the fractional potential which includes the first three terms, namely
$U_{3}$, whose graph is similar to the Newtonian one for $c_{2}>c_{1}/8$.

In Section 3 we present a zonal potential, which is of great help in modeling,
for example, the motion of a satellite moving in the equatorial plane of the Earth.

Section 4 is dedicated to the logarithmic Newtonian potential.

Section 5 is dedicated to the study of the existence and linear stability of circular orbits. We prove that circular orbits can be traced by a body moving in the field generated by the continued fractional potential $U_2$, and these orbits are linearly stable. 

In Section 6 we formulate some concluding remarks.

This qualitative study is useful because it offers a complete description of the potentials, for each value of the parameters involved; therefore in the following attempts to explain phenomena on various scales in the universe, the suitable potentials can be chosen knowing in advance their properties.

\section{Potentials generated by continued fractions}

We remind some definitions and properties concerning the continued fractions.
These can be found, for example, in the book of 
\cite{battin1999}, where the
author consider them as basic topics in analytical dynamics, and emphasize
their important role in many aspects of Astrodynamics.

Continued fractions were used at first to approximate irrational numbers, the
partial numerators and denominators being then integer numbers. A
\emph{continued fraction} is given by the expression%
\begin{equation}
\frac{a_{1}}{b_{1}+\dfrac{a_{2}}{b_{2}+\dfrac{a_{3}}{b_{3}+\ldots}}}%
:=\frac{a_{1}}{b_{1}+}\frac{a_{2}}{b_{2}+}\frac{a_{3}}{b_{3}+}\ldots\text{,}
\label{f1}%
\end{equation}
where the partial numerators $a_{n}$ ($n\in\mathbb{N}=\left\{  1,2,3,\ldots
\right\}  $) and the partial denominators $b_{n}$ ($n\in\mathbb{N}$) are real
(or complex) numbers.

An infinite sequence $(A_{n}/B_{n})_{n\in\mathbb{N}}$\ is associated to the
continued fraction (\ref{f1}) in the following way:%
\begin{equation}%
\begin{array}
[c]{l}%
\dfrac{A_{1}}{B_{1}}=\dfrac{a_{1}}{b_{1}},\ \ \dfrac{A_{2}}{B_{2}}%
=\dfrac{a_{1}}{b_{1}+}\dfrac{a_{2}}{b_{2}}=\dfrac{a_{1}b_{2}}{b_{1}b_{2}%
+a_{2}}\text{,}\\
\ \ \  \ldots\\
\dfrac{A_{n}}{B_{n}}=\dfrac{a_{1}}{b_{1}+}\dfrac{a_{2}}{b_{2}+}...\dfrac
{a_{n}}{b_{n}},\\
\ \ \  \ldots
\end{array}
\label{f2}%
\end{equation}
The fraction $A_{n}/B_{n}$\ is called a \emph{partial convergent} or simply a
\emph{convergent}. The expressions $A_{n}$\ and $B_{n}$\ satisfy the
fundamental recurrence formulas%
\begin{equation}%
\begin{array}
[c]{c}%
A_{n}=b_{n}A_{n-1}+a_{n}A_{n-2}\\
B_{n}=b_{n}B_{n-1}+a_{n}B_{n-2}%
\end{array}
,n=1,2,3\ldots\label{f3}%
\end{equation}
with initial conditions%
\begin{equation}
A_{-1}=1,A_{0}=0,B_{-1}=0,B_{0}=1. \label{f4}%
\end{equation}
This can be easily proved by mathematical induction. If the limit $\lim_{n\rightarrow
+\infty}A_{n}/B_{n}$ exists, it represents the \emph{value of the continued
fraction}; otherwise, the continued fraction is divergent.

Following \cite{battin1999} we mention some interesting monotonicity properties of
the sequence $A_{n}/B_{n}$, for $a_{n},b_{n}>0,$ $n\in\mathbb{N}$.

We calculate%
\[%
\begin{array}
[c]{l}%
\dfrac{A_{2n+1}}{B_{2n+1}}-\dfrac{A_{2n-1}}{B_{2n-1}}= -\dfrac{a_{1}a_{2}\ldots a_{2n+1}b_{2n+1}}{B_{2n+1}%
B_{2n-1}}<0\text{,}\\
\\
\dfrac{A_{2n}}{B_{2n}}-\dfrac{A_{2n-2}}{B_{2n-2}}=\dfrac{a_{1}a_{2}\ldots a_{2n}b_{2n}}{B_{2n}B_{2n-2}%
}>0\text{,}%
\end{array}
\]
and we remark that the odd convergents decrease, while the even convergents
increase. In a similar way, we calculate%
\[
\dfrac{A_{2n+1}}{B_{2n+1}}-\dfrac{A_{2n}}{B_{2n}}=\dfrac{a_{1}a_{2}\ldots a_{2n+1}}{B_{2n+1}B_{2n}}>0
\]
and it follows that every even convergent is smaller than every odd convergent.

We can summarize this as%
\begin{eqnarray}
\frac{A_{2}}{B_{2}}&<&\frac{A_{4}}{B_{4}}<\ldots<\frac{A_{2n}}{B_{2n}}%
<\ldots<\frac{A_{2n+1}}{B_{2n+1}}<\nonumber\\
&<&\ldots<\frac{A_{3}}{B_{3}}<\frac{A_{1}%
}{B_{1}}. \label{f4b}%
\end{eqnarray}

Using the idea of continued fractions, \cite{abd2014} considered a perturbation of the Newtonian potential. They started with the
continued fraction (\ref{f1}) and put for $r$, $c_{n}>0$, 
$n\in\mathbb{N}$,
\begin{eqnarray}\label{sirula}
r&=&b_{1}=b_{2}=\ldots, \\
a_{1}&=& \mu, \quad a_{2}=c_{1}\mu, \quad a_{3}=c_{2}\mu, \quad  \ldots,\nonumber 
\end{eqnarray}
to obtain%
\begin{equation}
\frac{\mu}{r+\dfrac{c_{1}\mu}{r+\dfrac{c_{2}\mu}{r+\ldots}}}:=\frac{\mu}%
{r+}\frac{c_{1}\mu}{r+}\frac{c_{2}\mu}{r+}\ldots\text{.}\label{f5}%
\end{equation}
Then, they retained only the first two terms
\[
\frac{\mu}{r+}\frac{c_{1}\mu}{r}=\frac{\mu}{r+\dfrac{c_{1}\mu}{r}}=\frac{\mu
r}{r^{2}+c_{1}\mu}%
\]
in order to obtain the second convergent, and got the potential
\begin{equation}
\frac{\mu r}{r^{2}+c_{1}\mu},\label{f6}%
\end{equation}
called \emph{continued fractional potential} or simply \emph{fractional
potential}. In this formula, when applied to the two-body problem, $r$\ stands for the mutual distance between two
punctual bodies and\ $\mu$ for the product of the gravitational constant
$G$\ with the sum of masses $m_{1}$\ and $m_{2}$. In what follows we shall
consider $c_{1},$ $c_{2},$ $\ldots>0$.

We remark that it is unnecessary and physically unsustained to put the constant $\mu$ everywhere in the continuous fraction. In what follows we shall consider 
a simplified form by mentaining $r=b_{1}=b_{2}=\ldots$, as in (\ref{sirula}), but changing $a_n$ into
$$ 
a_{1}= \mu, \quad a_{2}=c_{1}, \quad a_{3}=c_{2}, \quad \ldots. 
$$

We obtain a simplified form of (\ref{f5}), namely 
\begin{equation}\label{f511}
\frac{\mu}{r+\dfrac{c_{1}}{r+\dfrac{c_{2}}{r+\ldots}}}:=\frac{\mu}%
{r+}\frac{c_{1}}{r+}\frac{c_{2}}{r+}\ldots\text{.}
\end{equation}

It is obvious that from (\ref{f511}) we get as the first convergent the
Newtonian potential
\begin{equation}
U_{1}(r)=\frac{\mu}{r}, \label{f7}%
\end{equation}
by keeping only the first term, and as a second convergent 
\begin{equation}
U_2(r)=\frac{\mu r}{r^{2}+c_{1}}.\label{f611}
\end{equation}
Now, using the first three terms, we obtain the potential
\begin{equation}
U_{3}(r)=\frac{\mu(r^{2}+c_{2})}{r\left(r^{2}+c_{1}+c_{2}\right)}.
\label{f8}%
\end{equation}
Similarly, using the first four terms, we get the potential%
\begin{equation}
U_{4}(r)=\frac{\mu r(r^{2}+c_{2}+c_{3})}{r^{4}+(c_{1}+c_{2}+c_{3})
r^{2}+c_{1}c_{3}} \label{f9}%
\end{equation}
and so on.

It is known that the Newtonian potential (\ref{f7}), regarded as a function of
$r$ defined on $(0,+\infty)$,\ is decreasing from $+\infty$\ to $0$. Therefore, it is natural to study the monotonicity of the other fractional
potentials and their limits at $0$\ and $+\infty$.

We begin with the behaviour of the continued fractional potential
$U_{2}:[0,+\infty)\rightarrow\lbrack0,+\infty)$, which has $U_{2}(0)=0$\ and
$\lim\nolimits_{r\rightarrow+\infty}U_{2}(r)=0$. 
The first two derivatives of $U_{2}$ are respectively
\begin{equation}
U_{2}^{\prime}(r)=\frac{\mu(-r^{2}+c_{1})}{(r^{2}+c_{1})^{2}}, U_{2}^{\prime\prime}(r)=-\frac{2\mu r(-r^{2}+3c_{1})}{(r^{2}+c_{1})^{3}}. \label{f10}
\end{equation}
The first derivative of $U_{2}$ has a unique positive root $\sqrt{c_{1}}$, and the second derivative of $U_{2}$ has a root equal to $0$ and a
positive root $\sqrt{3c_{1}}$. It follows that the potential $U_{2}%
$\ increases from $0$\ to $\sqrt{1/c_{1}}/2=U_{2}\left(  \sqrt{c_{1}
}\right)  $ and then it decreases to $0$\ as $r\rightarrow+\infty$, having an
inflection point at $\sqrt{3c_{1}}$. So, whatever the positive value of
the constant $c_{1}$\ is, the graph of the continued fractional potential
$U_{2}$\ looks different from that of the Newtonian potential for
$r$\ relatively small. The dot graph from Fig. 3 represents $U_{2}$\ for
$c_{1}=0.1$\ and $\mu=1$.

We consider now the continued fractional potential $U_{3}:(0,+\infty
)\rightarrow(0,+\infty)$, for which $\lim\nolimits_{r\rightarrow0+}%
U_{3}(r)=+\infty$ and $\lim\nolimits_{r\rightarrow+\infty}U_{3}(r)=0$. The
first two derivatives of $U_{3}$\ are respectively%
\begin{equation}
\begin{split}
U_{3}^{\prime}(r)&=-\dfrac{\mu(r^{4}-(c_{1}-2c_{2})r^{2}+c_{2}%
(c_{1}+c_{2}))}{r^{2}\left(  r^{2}+c_{1}+c_{2}\right)^{2}},\\
U_{3}^{\prime\prime}(r)&=\dfrac{2\mu(r^{6}-3(c_{1}-c_{2}) r^{4}+3c_{2}%
(c_{1}+c_{2})r^{2})}{r^{3}(r^{2}+c_{1}+c_{2})^{3}}\\
&+\dfrac{2c_2\mu(c_{1}+c_{2})^{2}}{r^{3}(r^{2}+c_{1}+c_{2})}.
\end{split}\label{f11}
\end{equation}

In order to study the monotonicity of $U_{3},$\ we denote $r^{2}=u$\ and
consider the numerator of $U_{3}^{\prime}$. The equation%
\begin{equation}
u^{2}-(c_{1}-2c_{2})u+c_{2}(c_{1}+c_{2})=0 \label{f12}%
\end{equation}
has the discriminant $\Delta=c_{1}(c_{1}-8c_{2})$. For $c_{2}>c_{1}%
/8$, the equation (\ref{f12})\ has no real roots, hence $U_{3}^{\prime}(r)<0$;
then $U_{3}(r)$\ strictly decreases from $+\infty$\ to $0$. For $c_{2}%
=c_{1}/8$, the equation (\ref{f12})\ has a double real root $u_{1,2}=3c_{1}%
/8$, hence $U_{3}^{\prime}(r)\leq0$\ and $U_{3}(r)$\ strictly decreases from
$+\infty$\ to $0$, and its graph has an inflection point at $r=\sqrt{3c_{1}%
/8}$ (where $U_{3}^{\prime\prime}(\sqrt{3c_{1}/8})=0$). The last case is
$c_{2}<c_{1}/8$, when (\ref{f12}), hence $U_{3}^{\prime}(r)$\ also, has two
distinct positive roots. In this situation, $U_{3}(r)$\ strictly decreases
from $+\infty$\ to a local minimum, then it strictly increases to a local
maximum, and finally decreases to $0$.

We illustrate in Fig. 1 these cases for $U_{3}$\ with\ $\mu=1$ and

a) $c_{1}=0.3$, $c_{2}=0.1$ (dash);

b) $c_{1}=0.8$, $c_{2}=0.1$ (dot);

c) $c_{1}=1$, $c_{2}=0.1$ (solid).%
\begin{figure}[ptb]%
\centering
\includegraphics[scale=0.5]%
{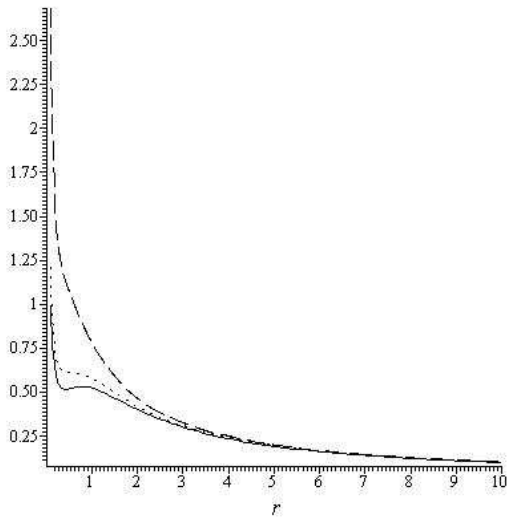}
\caption{$U_{3}$ for $c_{1}=0.3;\ 0.8;\ 1$ and $c_{2}=0.1,\ \mu=1$}%
\end{figure}

In conclusion, the continued fractional potential $U_{3}$\ has three kinds of
graphs, depending on the values of the coefficients $c_{1}$\ and $c_{2}$. The
graph of $U_{3}$\ is similar to that of the Newtonian potential for
$c_{2}>c_{1}/8$. We emphasize that the behaviour of $U_{3}$\ in the
vicinity of $0$, for any $c_{1}$, $c_{2}>0$,\ is similar to that of the
Newtonian potential, in contrast with that of $U_{2}$ given in (\ref{f8}).

It follows that a good choice for a continued fractional potential, which is
close to the Newtonian one and still easy to handle, is $U_{3}$ with
$c_{2}>c_{1}/8$.

We consider the continued fractional potential $U_3$ with $c_1=0.0001$ and $c_2=0.00002$, and we apply it for {\it Raduga-1M 2} GEO satellite with semi-major axis 42164 km and the period of revolution 1436 minutes. The orbit is displayed in Fig. 2. This choice of the small parameters $c_1$ and $c_2$ is situated in the range $c_2 > c_1/8$, so that the potential $U_3$ is very similar to the Newtonian one. 
\begin{figure}[ptb]%
\centering
\includegraphics[scale=0.4]%
{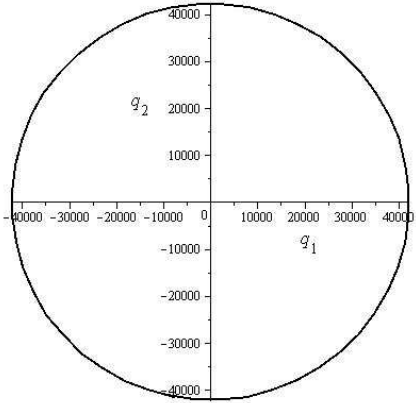}%
\vspace{0.6cm}
\caption{{\it Raduga-1M 2} satellite orbit using continued fractional potential  $U_{3}$, $r=\sqrt{q_1^2+q_2^2}$}%
\end{figure}

It can be easily checked that%
\begin{equation}
U_{1}(r)-U_{3}(r)>0, \quad U_{4}(r)-U_{2}(r) >0.
\end{equation}
In fact, from relation (\ref{f4b}) we obtain that in general%
\begin{eqnarray}
U_{2}(r)<U_{4}(r)<\ldots<U_{2n}(r)<\ldots<U_{2n+1}(r)< & & \nonumber\\
\ldots<U_{3}%
(r)<U_{1}(r). & &  \label{f13}%
\end{eqnarray}

 In Fig. 3 we plot the graphs of the first four continued
fractional potentials $U_{1}$ (dash), $U_{2}$ (dot), $U_{3}$ (dashdot), and
$U_{4}$ (solid) for $\mu=1,$ $c_{1}=c_{2}=c_{3}=0.1$.\
\begin{figure}[ptb]%
\centering
\includegraphics[scale=0.5]%
{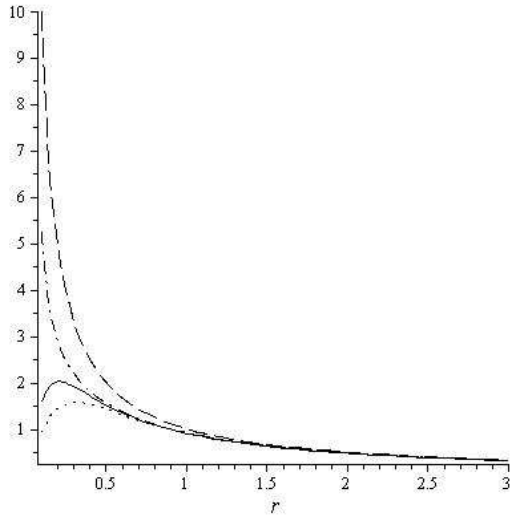}
\caption{$U_{1},\ U_{2},\ U_{3},\ U_{4}$ for $c_{1}=c_{2}=c_{3}=0.1$ and
$\mu=1$}%
\end{figure}

\section{On the zonal potential of Earth}

Until now we considered the motion of two punctual masses $m_{1}$\ and $m_{2}$. In the case of the very important problem of the motion of artificial
satellites around the Earth, the Earth cannot be approximated by a point in
order to obtain a good model of a satellite's motion; a good enough
approximation is that of a spheroid.

It is known \citep{king1963,roy2005} that if it is
assumed that the Earth is a spheroid (i. e. we neglect the tesseral and
sectorial harmonics), then its potential may be written as a series of zonal
harmonics of the form%
\begin{equation}
U_{zon}=\frac{\mu}{r}\left(  1-\sum\limits_{n=2}^{\infty}J_{n}\left(  \frac
{R}{r}\right)  ^{n}P_{n}(\sin\phi)\right)  \text{,} \label{f40}%
\end{equation}
where $\mu=GM$\ is the product of the gravitational constant $G$\ with the
mass of the Earth $M$,\ $R$\ is the equatorial radius of the Earth and $J_{n}%
$\ are the zonal harmonic coefficients due to the oblateness of the Earth. The
coordinates of the satellite are the distance to the center of the Earth
$r$\ and its latitude $\phi$,\ and $P_{n}(\cdot)$ are the Legendre polynomials
of degree $n$.\ Equation (\ref{f40}) does not take into account the small
variation of $U_{zon}$\ with longitude. Because the coefficients $J_{n}%
$,\ $n>2$,\ are much smaller than $J_{2}$,\ a good approximation of the zonal
potential of Earth is%
\begin{equation}
U_{zonal}=\frac{\mu}{r}\left[  1-\frac{1}{2}J_{2}\left(  \frac{R}{r}\right)
^{2}(3\sin^{2}\phi-1)\right]  \text{.} \label{f50}%
\end{equation}

In order to compare it with the Newtonian and continued fractional potentials,
we consider equatorial orbits, with $\phi=0$. The potential is then of the
type%
\begin{equation}
U=\frac{\mu}{r}+\frac{\mu c}{r^{3}},\ \ c>0, \label{f51}%
\end{equation}
with $c=R^{2}J_{2}/2$. 

We remark that $U$ is an inhomogeneous potential. Such potentials, in relation
with the families of orbits generated by them, are studied by \cite{bozis1997}.

The potential $U$ has $\lim\nolimits_{r\rightarrow0+}U(r)=+\infty$ and
$\lim\nolimits_{r\rightarrow+\infty}U(r)=0$. The first two derivatives are
respectively%
\begin{equation}
U^{\prime}(r)=-\frac{\mu(r^2+3c)}{r^{4}},\ \ \ U^{\prime\prime}(r)=\frac
{2\mu(r^2+6c)}{r^{5}}. \label{f52}%
\end{equation}
The first derivative of $U$\ is negative and its second derivative is positive
on $(0,+\infty)$, hence the potential $U$\ decreases from its limit in $0$,
which is $+\infty$,\ to its limit $0$\ as $r\rightarrow+\infty$. Therefore,
the zonal potential $U$\ has a graph similar to the Newtonian one, as it may
be seen in Fig. 4, for $\mu=1$ and $c=1$.%
\begin{figure}[ptb]
\centering
\includegraphics[scale=0.5]{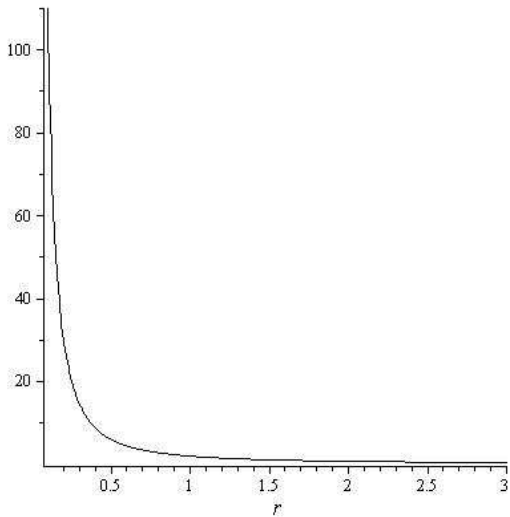}
\vspace{0.6cm}
\caption{The zonal potential $U$ for $c=1$ and $\mu=1$}%
\end{figure}
For the sake of completeness, we study also the case of $c<0$. The first
derivative of $U$\ has now the positive root $\sqrt{-3c}$,\ and the second derivative
of $V$\ has the positive root $\sqrt{-6c}$. It follows that the potential
$U$\ increases from its limit in $0$, which is equal with $-\infty$\ to
$U\left(  \sqrt{-3c}\right)  >0$ and then it decreases to $0$\ as
$r\rightarrow+\infty$, having an inflection point at $\sqrt{-6c}$.\ So, whatever the
negative value of the constant $c$\ is, the graph of the zonal-type potential
$V$\ looks different from that of the Newtonian potential.

The fact that the zonal potential $U$,\ for positive $c$,\ has the graph
similar to that of the Newtonian potential, while the fractional potential
$U_{2}$\ has the dotted graph from Fig. 3, explains difference of the
corresponding plotted graphs for low and medium altitudes in Figs. 4 - 5 of \cite{abd2014}.

Applying the zonal potential $U$ for the {\it Raduga-1M 2} satellite motion, we obtain the trajectory displayed in Fig. 5.
\begin{figure}[ptb]%
\centering
\includegraphics[scale=0.4]%
{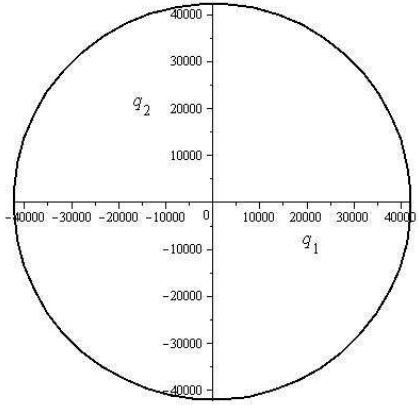}%
\vspace{0.6cm}
\caption{{\it Raduga-1M 2} satellite motion using the zonal potential $U$, $r=\sqrt{q_1^2+q_2^2}$}%
\end{figure}

\section{Logarithmically corrected potential}

\cite{fabries2009} have analyzed the rotation curves of some
spiral galaxies, using a disc modelization, with a Newtonian potential
corrected with an extra logarithmic term. More recently, 
\cite{ragos2013} have taken into account the effects in the anomalistic
period of celestial bodies due to the same logarithmic correction to the
Newtonian gravitational potential. We shall compare this corrected potential
with the Newtonian one. The corrected potential $V:(0,+\infty)\rightarrow
(0,+\infty)$ is given by%
\begin{equation}
V(r)=\frac{\mu}{r}+\mu\alpha\log\left(  r\right)  \text{.} \label{f100}%
\end{equation}
Its first two derivatives are respectively%
\begin{equation}
V^{\prime}(r)=\frac{\mu(\alpha r-1)}{r^{2}}\text{,}\ \ \ V^{\prime\prime
}(r)=-\frac{\mu(\alpha r-2)}{r^{3}}\text{.} \label{f101}%
\end{equation}

Let us consider that the coefficient $\alpha$\ is positive. Then
$\lim\nolimits_{r\rightarrow0+}V(r)=+\infty$ and $\lim\nolimits_{r\rightarrow
+\infty}V(r)=+\infty$. The first derivative of $V$\ has the positive root
$1/\alpha$,\ and the second derivative of $V$\ has the positive root
$2/\alpha$. It follows that the potential $V$\ decreases from its limit in
$0$, which is equal with $+\infty$\ to $\alpha(1-\log\left(  \alpha\right)
)=V\left(  1/\alpha\right)  $ and then it increases to $+\infty$\ as
$r\rightarrow+\infty$, having an inflection point at $2/\alpha$.\ So, whatever
the positive value of the constant $\alpha$\ is, the graph of the logarithmically corrected potential $V$\ looks different from that of the Newtonian potential. We illustrate in Fig. 6 the shape of $V$\ for $\mu=\alpha=1$.%
\begin{figure}[ptb]%
\centering
\includegraphics[scale=0.5]%
{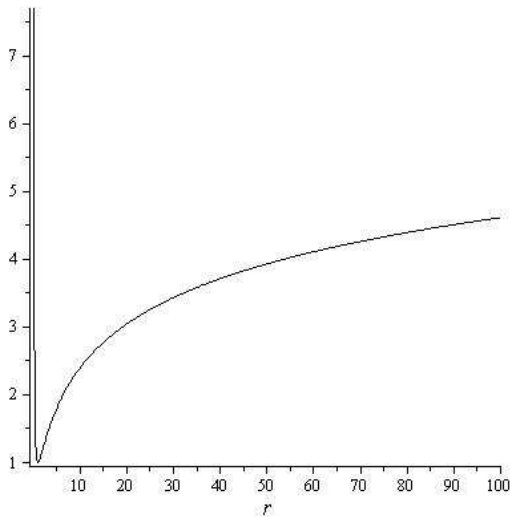}%
\caption{The logarithmically corrected potential $V$\ for $\mu=\alpha=1$}%
\end{figure}
For a negative $\alpha$,\ we have $\lim\nolimits_{r\rightarrow0+}V(r)=+\infty
$, but $\lim\nolimits_{r\rightarrow+\infty}V(r)=-\infty$. The first derivative
of $V$\ is negative and its second derivative is positive on $(0,+\infty)$,
hence the potential $V$\ decreases from its limit in $0$, which is also equal
with $+\infty$,\ to its limit $-\infty$\ as $r\rightarrow+\infty$. We
represent in Fig. 7 the potential $V$\ for $\mu=1$ and $\alpha=-1$.%
\begin{figure}[htb]%
\centering
\includegraphics[scale=0.5]%
{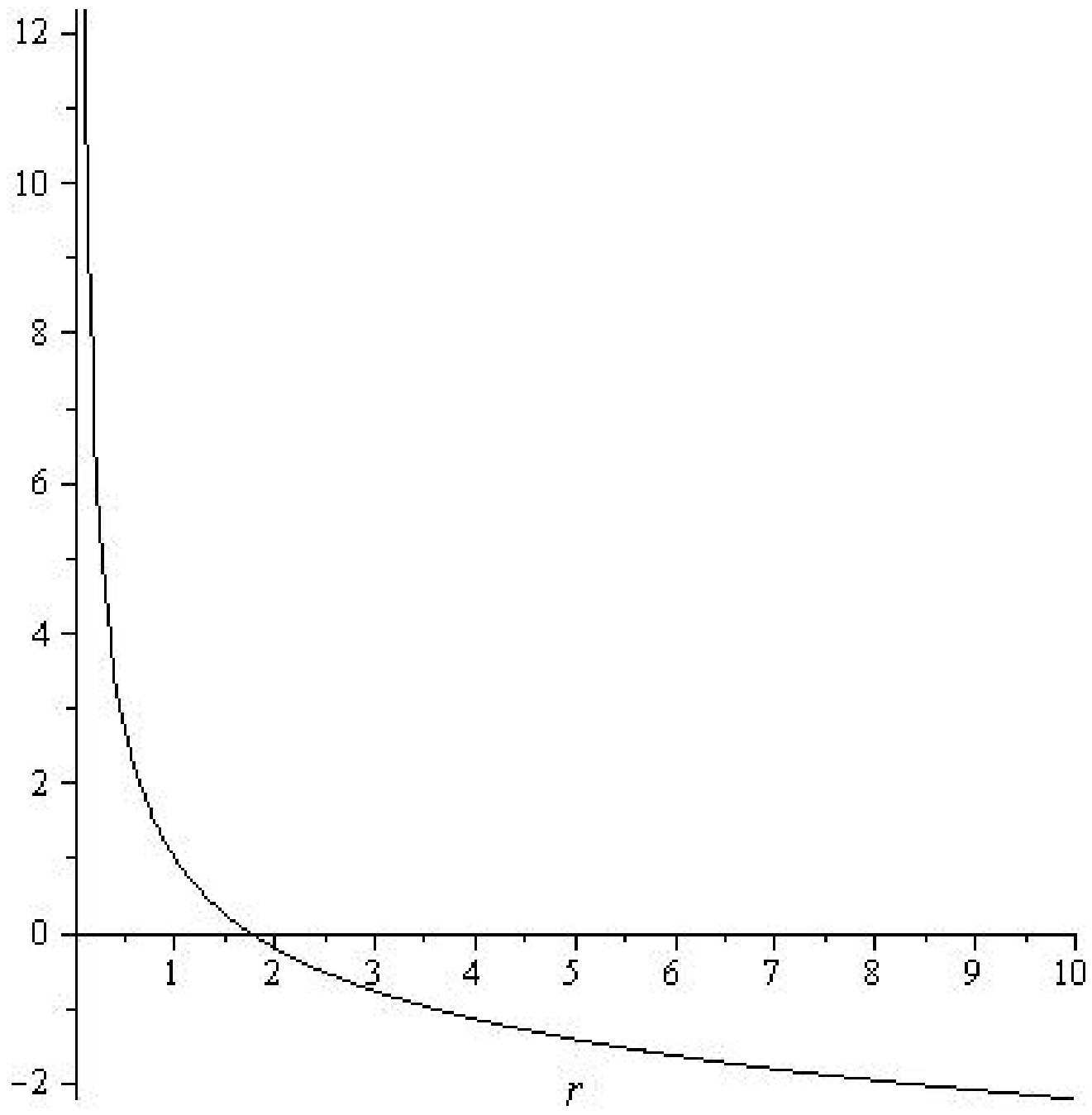}%
\caption{The logarithmically corrected potential $V$ for $\mu=1$ and $\alpha=-1$}%
\end{figure}
We consider the logaritmically corrected potential $V$ with $\alpha=-0.0001$, and we apply it for the motion of a star from the dynamical system like "Milky-way" spiral galaxy with initial position at 8 kpc, initial velocity 572 kpc/Gyr and the period of revolution 0.04 Gyr. The orbit is displayed in Fig. 8.
\begin{figure}[htb]%
\centering
\includegraphics[scale=0.4]%
{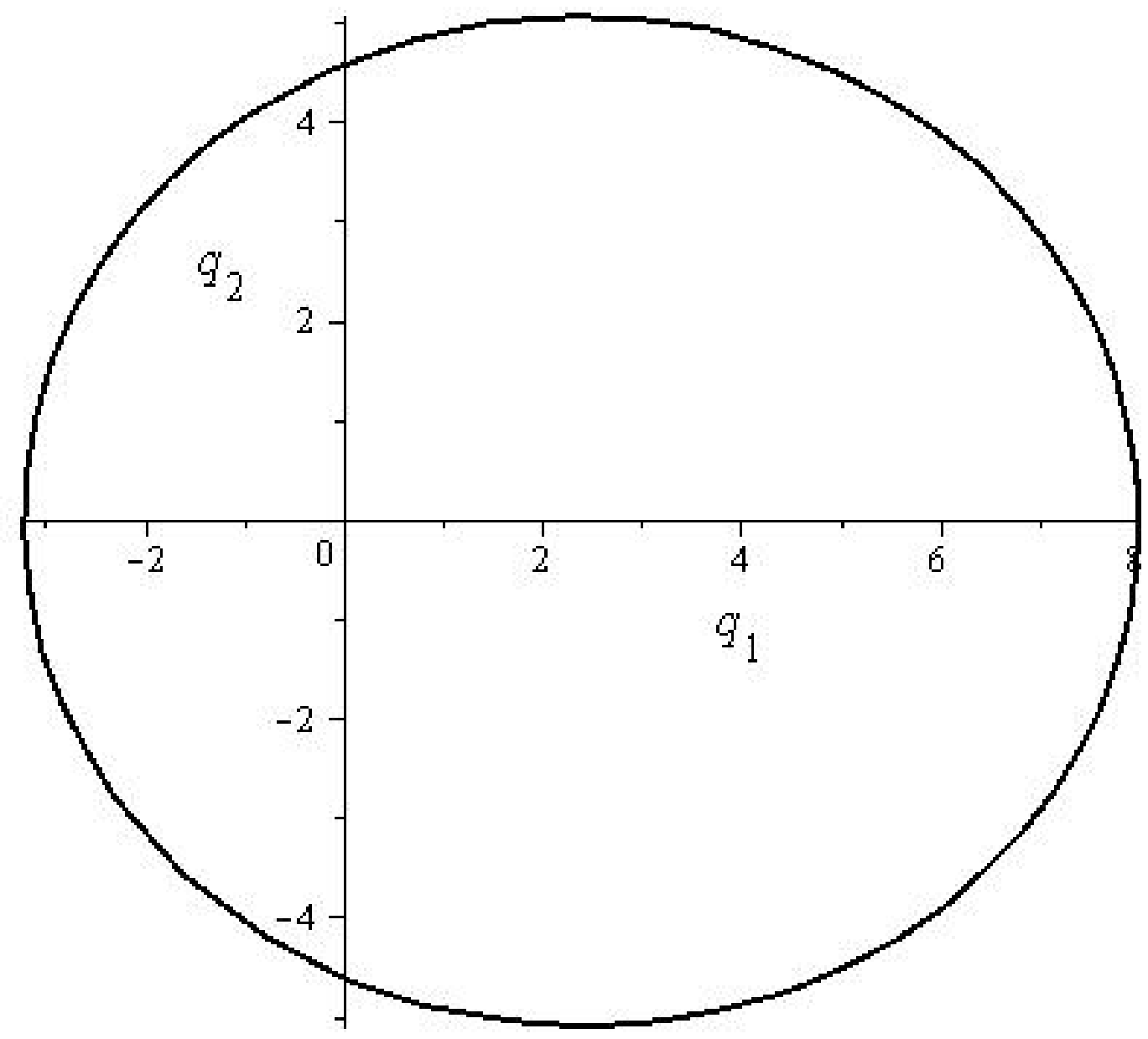}%
\vspace{0.6cm}
\caption{Motion of a star in a dynamical system like "Milky way" spiral galaxy
using the logarithmically corrected potential $V$ for $\alpha=-0.0001$}%
\end{figure}

\section{Circular orbits in the continued fractional potential $U_{2}$}

Circular orbits appear in the motion of the equatorial satellites, of some
planets in various planetary systems, or of stars in galaxies. Therefore it is  important to study if they can be traced in the continued fractional fields.
We shall prove
that such orbits can be traced by a body moving in the field produced by the
continued fractional potential $U_{2}$, which is a generalization of the
Newtonian potential. The existence and the stability of circular orbits in a
Maneff field was studied recently by \cite{b}.

We consider a two-body problem with a primary body of mass $M$ and a secondary
one of mass $m$, under the influence of the continued fractional potential
$U_{2}$. The potential is central, so the two-body problem may be reduced to a
central force one, and we shall study the relative motion of the secondary
body. The motion is planar and it is governed, in polar coordinates
$r=\sqrt{q_{1}^{2}+q_{2}^{2}}$, $\theta=\tan^{-1}(q_{2}/q_{1})$, by the
equations%
\begin{equation}%
\begin{tabular}
[c]{l}%
$\ddot{r}-r\dot{\theta}^{2}=U_{2}^{\prime}(r)$\\
$\dfrac{\operatorname*{d}}{\operatorname*{d}t}(r^{2}\dot{\theta})=0,$%
\end{tabular}
\ \label{aa1}%
\end{equation}
where $U_{2}^{\prime}$\ is given by (14) with $\mu=GM$ and $c_{1}$\ is a
positive constant.

From the second equation we get
\begin{equation}
r^{2}\dot{\theta}=h,\label{aa2}%
\end{equation}
where $h$\ is a constant.

For $h\neq 0$, we try for a constant solution $r=r_{0}$, $r_{0}\neq0$ for the
equations (\ref{aa1}). For such a solution, equation (\ref{aa2}) gives
$\dot{\theta}=h/r_{0}^{2}=\omega_{0}$. The first equation of (\ref{aa1}) reads
for $r(t)=r_{0}$:%
\begin{equation}
h^{2}+r_{0}^{3}U_{2}^{\prime}(r)|_{r=r_{0}}=0.\label{aa3}%
\end{equation}

If this equation admits a solution $r_{0}>0$, it means that a circular orbit
$r(t)=r_{0}$\ is possible, the secondary pursuing the orbit with constant
angular velocity $\omega_{0}=h/r_{0}^{2}$.

Such a circular orbit is linearly stable if%
\begin{equation}
U_{2}^{\prime\prime}(r)|_{r=r_{0}}+\frac{3}{r_{0}}U_{2}^{\prime}(r)|_{r=r_{0}%
}<0, \label{aa4}%
\end{equation}
this condition being obtained by developing $U_{2}$\ around $r_{0}$\ up to the
first-order terms \citep{w,roy2005}.

By applying this reasoning for the Newtonian potential given by (10), with
$\mu>0$, we obtain easily that the circular orbit $r_{N}=h^{2}/\mu$\ obtained
from the equation corresponding to (\ref{aa3}) is linearly stable, since%
\[
U_{1}^{\prime\prime}(r)|_{r=r_{N}}+\frac{3}{r_{0}}U_{1}^{\prime}(r)|_{r=r_{N}%
}=-\frac{\mu}{r_{N}^{3}}<0.
\]

We study now the case of the continued fractional potential $U_{2}$ given by
(11), with $\mu,c_{1}>0$. Equation (\ref{aa3}), where we denote shortly
$r=r_{0}$, reads%
\[
h^{2}+r^{3}\frac{\mu(c_{1}-r^{2})}{(r^{2}+c_{1})^{2}}<0.
\]
This is equivalent with%
\begin{equation}
-\mu r^{5}+h^{2}r^{4}+\mu c_{1}r^{3}+2c_{1}h^{2}r^{2}+h^{2}c_{1}%
^{2}=0\label{aa5}%
\end{equation}
and this fifth degree equation cannot be solved in general. Nevertheless, we
remark that all the coefficients, excepting that of $r^{5}$,\ are positive. It
follows that there is precisely one change of sign in the row of the
coefficients. Applying the Descartes rule of signs \citep{kk} it
follows that equation (\ref{aa5}) has at most one positive solution. But for
$r=0$\ the left hand side of (\ref{aa5}) is equal to $h^{2}c_{1}^{2}>0$,\ and
its limit to infinity is $-\infty$, hence equation (\ref{aa5}) has a unique
solution $r=r_{0}$. Moreover, for $r_{N}=h^{2}/\mu$\ the left hand side of
(\ref{aa5}) is positive, which means that $r_{0}>r_{N}$, i. e. the radius of
the circular orbit in the fractional potential $U_{2}$\ is greater than the
similar one traced in the Newtonian potential.

To get information on the stability of the unique circular orbit $r=r_{0}$, we
calculate the left hand side of (\ref{aa4}), using the expressions of the
derivatives of $U_{2}$\ from (14):%
\begin{equation}
U_{2}^{\prime\prime}(r)+\frac{3}{r}U_{2}^{\prime}(r)=-\mu\frac{r^{4}%
+6c_{1}r^{2}-3c_{1}^{2}}{r(r^{2}+c_{1})^{3}}.\label{aa6}%
\end{equation}
The sign of this expression is given by the sign of%
\[
g(r)=-r^{4}-6c_{1}r^{2}+3c_{1}^{2}.
\]
This polynomial of fourth degree in $r$\ has a unique positive root%
\begin{equation}
r_{1}=\sqrt{c_{1}(2\sqrt{3}-3)},\label{aa7}%
\end{equation}
hence it has positive values on the interval $(0,r_{1})$\ and negative ones on
$(r_{1},+\infty)$.

We have proved that the fifth degree polynomial in (\ref{aa5}) is positive on
$(0,r_{0})$\ and negative on $(r_{0},+\infty)$. We calculate the value of that
polynomial for $r_{1}$\ given by (\ref{aa7})\ and get:%
\[%
\begin{tabular}
[c]{l}%
$\mu r^3_{1}(c_{1}-r_{1}^{2})+h^{2}(r_{1}^{4}+2c_{1}r_{1}^{2}+c_{1}^{2})=$\\
$\mu r_{1}^{3}(c_{1}-c_{1}(2\sqrt{3}-3))+h^{2}(r_{1}^{2}+c_{1})^{2}=$\\
$2(2-\sqrt{3})\mu c_{1}r_{1}^{3}+h^{2}(r_{1}^{2}+c_{1})^{2}>0.$%
\end{tabular}
\
\]
It follows that $r_{1}<r_{0}$, hence $r_{0}\in(r_{1},+\infty)$\ and
$g(r_{0})<0$. Then%
\[
U_{2}^{\prime\prime}(r_{0})+\frac{3}{r}U_{2}^{\prime}(r_{0})<0
\]
and the orbit $r=r_{0}$\ is linearly stable.

\section{Conclusion}

The main properties of the Newtonian potential are preserved by some of its
corrected potentials: the continued fractional potential $U_{3}$ given by
(\ref{f8}), for $c_{2}>c_{1}/8>0$; the zonal potential of Earth $U$ given by
(\ref{f51}). It is worth noting that in the case of the
continued fraction potential $U_{3}$ we have at our disposal two parameters
$c_{1}$ and $c_{2}$. These can be used to adjust the potential when we have information on the motion of the satellite. The figures illustrate the possible situations which have been proved analytically.

For the continued fractional potential $U_{2}$ it is proved that circular
orbits exist and are linearly stable.

We remark a strong feature of the continued fractional potentials: they have a simple analytical form, being rational functions, hence they can be easily handled in further applications. 


\acknowledgments
The authors are deeply indebted to the reviewers and to the editor for their valuable comments and suggestions.

The work of the second author was partially supported by a grant of the Romanian National Authority for Scientific Research, CNDI-UEFISCDI, project number PN-II-PT-PCCA-2011-3.2-0651 (AMHEOS).

\end{document}